\documentclass[twocolumn,showpacs,preprintnumbers,amsmath,amssymb]{revtex4}

\usepackage{graphicx}
\usepackage{dcolumn}
\usepackage{bm}

\newcommand{\fex}{f^{\rm ex}} 
\newcommand{\fideal}{f^{\rm id}} 
\newcommand{\eq}[1]{~(\ref{#1})} 

\newcommand{\be}{\begin{equation}}
\newcommand{\ee}{\end{equation}}

\newcommand{\rh}{\rho}  
\newcommand{\sig}{\sigma} 
\newcommand{\pol}{\delta} 

\newcommand{\muex}{\mu^{\rm ex}} 

\newcommand{\fexhs}{\fex_{\rm hs}} 

\newcommand{\rat}{\xi} 
\newcommand{\ratc}{\xi_{\rm c}} 
\newcommand{\sigp}{\sig_{\rm p}} 
\newcommand{\sigc}{\sig_{\rm c}} 

\newcommand{\mup}{\mu_{\rm p}} 
\newcommand{\mupsig}{\mup(\sigp)} 
\newcommand{\mupex}{\mu_{\rm p}^{\rm ex}(\sigp)} 

\newcommand{\rhpsig}{\rh_{\rm p}(\sigp)} 
\newcommand{\rhcsig}{\rh_{\rm c}(\sigc)} 

\newcommand{\rhp}{\rh_{\rm p}} 
\newcommand{\rhc}{\rh_{\rm c}} 

\newcommand{\mpi}{\rhp{}_i} 

\newcommand{\phic}{\phi_{\rm c}}
\newcommand{\phip}{\phi_{\rm p}}

\newcommand{\hs}{_{\rm hs}}

\newcommand{\rats}{\rat_S}
\newcommand{\ratn}{\rat_N}
\newcommand{\ratm}{\rat_M}
\newcommand{\ratb}{\rat_B}

\newcommand{\np}{n_{\rm p}}

\begin{document}


\title{Effects of polymer polydispersity on the phase behaviour of
colloid-polymer mixtures}
       
\author{Moreno Fasolo}
\author{Peter Sollich}
\email{peter.sollich@kcl.ac.uk}
\affiliation{
Department of Mathematics, King's College London, London WC2R 2LS, U.K.
}

\date{\today}

\begin{abstract}
We study the equilibrium behaviour of a mixture of monodisperse hard
sphere colloids and polydisperse non-adsorbing polymers at their
$\theta$-point, using the Asakura-Oosawa model treated within the
free-volume approximation.
Our focus is the experimentally relevant scenario where the
distribution of polymer chain lengths across the system is fixed.
Phase diagrams are calculated using the moment free energy method, and
we show that the mean polymer size $\ratc$ at which gas-liquid phase
separation first occurs decreases with increasing polymer
polydispersity $\pol$. Correspondingly, at fixed mean polymer size,
polydispersity favours gas-liquid coexistence but delays the onset of
fluid-solid separation. On the other hand, we find that systems with
different $\pol$ but the same {\em mass-averaged} polymer chain length
have nearly polydispersity-independent phase diagrams. We conclude
with a comparison to previous calculations for a semi-grandcanonical
scenario, where the polymer chemical potentials are imposed; there it
was found that fluid-solid coexistence was favoured over gas-liquid in
some areas 
of the phase diagram. Our results show that this somewhat
counter-intuitive result arose because the actual polymer size
distribution in the system is shifted to smaller sizes relative to the
polymer reservoir distribution.
\end{abstract}

\maketitle

\section{Introduction}
	

Colloid-polymer mixtures have a wide variety of practical uses:
polymers can be added in low concentrations to colloidal suspensions
to modify their properties~\cite{Vincent74,Napper83,RusSavSch89}, or
conversely colloids can be added to polymer melts as
``fillers''~\cite{EniFriStaPop90} or to polymer solutions to produce
gels~\cite{Lewis00}. In the former case, it is well known
that the presence of the polymer can induce an effective
colloid-colloid {\em depletion
attraction}~\cite{AsaOos54,AsaOos58,Vrij76,JoaLeiDeG79}.  Unlike in
atomic systems, both the strength and range of this interaction can be
tuned by varying polymer concentration and size. The implications of
this for the phase behaviour of the mixture are of fundamental
theoretical interest, and have been the subject of much research.

We focus on the simplest situation in this paper, where the colloids
have hard interactions with each other and with the polymers; this
requires in particular that the polymers be non-adsorbing on the
surface of the colloids. We further restrict ourselves to
$\theta$-point conditions, where polymer-polymer interactions can be
neglected to a first approximation. When the polymers are modelled as
random walks on a lattice, this scenario is amenable to direct
numerical simulation~\cite{MeiFre91,MeiFre94}. However, most work has
focussed on a further simplification, proposed by Asakura and Oosawa
(AO)~\cite{AsaOos54,AsaOos58} and Vrij~\cite{Vrij76}, where the
polymers are replaced by spheres with radius equal to the radius of
gyration of the original polymer chains. These ``polymer spheres'' are
then assumed to interpenetrate freely with each other while having a
hard excluded volume interaction with the colloids. This AO model has
been studied in great detail, by thermodynamic perturbation
theory~\cite{GasHalRus83,Vincent87,VinEdwEmmCro88}, free-volume
theory~\cite{LekPooPusStrWar92} (see below), density functional
theory~\cite{SchLowBraEva00,SchLowBraEva02,BraEvaSch03}, and Monte
Carlo simulations~\cite{MeiFre94,DijBraEva99,BolLouHan02}. Note
that the naive choice of setting the radius of the effective polymer
sphere equal to the polymer radius of gyration can be improved upon;
we will return to this point in the discussion in Sec.~\ref{sec:conclusion}.

In addition to the simplifications already discussed, the AO model
ignores the fact that there is, in real colloid-polymer mixtures,
invariably an (essentially continuous) spread of polymer chain
lengths. The effects of such {\em polymer size polydispersity} on the
phase behaviour have been studied by some researchers. As described in
more detail below, however, the experimentally relevant situation
where the full polymer size distribution in the system is conserved
during phase separation remains to be understood. This is the issue
which we address in this paper. Specifically, we study the AO model
within the free-volume approximation of~\cite{LekPooPusStrWar92}, for
the case where the colloids are monodisperse, with diameter $\sigc$,
while the polymer spheres are polydisperse with density distribution
$\rhpsig=\rhp \np(\sigp)$. Here $\rhp$ is the polymer number density
and $\np(\sigp)$ is the normalised polymer diameter distribution. We
take the colloid diameters $\sigc$ as our unit of length so that the
polymer-to-colloid diameter ratio $\rat$ coincides with the
dimensionless polymer diameter, $\rat\equiv\sigp$. All densities are
made dimensionless by multiplying by the volume of a colloid particle,
and quantities with the dimension of energy are measured in units of
$k_{\rm B}T$.
%

For the case of monodisperse polymers, it is well known that the phase
behaviour depends strongly on the polymer-to-colloid diameter ratio
(see e.g.~\cite{FasSol04b,LekPooPusStrWar92}). For small polymers only
coexistence between colloidal fluid and solid phases is observed.
(The term ``fluid'' is used here because there are no distinguishable
gas and liquid phases in such a system.)  For larger polymers, on the
other hand, with $\rat$ above some threshold value $\ratc$, a region
of gas-liquid phase separation can occur in the phase diagram. When
generalising these considerations to the case of polydisperse
polymers, there is no longer a single polymer size but rather a
continuum of different diameters $\sigp$. A choice therefore needs to
be made on how to define appropriately the typical polymer size that
determines the phase behaviour. Ideally, one would like this
definition to be such that phase diagrams become nearly independent of
polymer polydispersity and can therefore be inferred from the
corresponding monodisperse reference system~\cite{Warren97}.

A naive definition of the typical polymer size is the mean polymer
diameter, $\rats=\int\!d\sigp\,\sigp \np(\sigp)$. This is, in fact,
independent of the diameter polydispersity $\pol$, defined as the
standard deviation of $\np(\sigp)$ normalised by its mean.  We will
see below, however, that this provides a poor mapping from
polydisperse to monodisperse systems, with phase behaviour at fixed
$\rats$ strongly dependent on $\pol$. Following~\cite{Warren97}, we
have therefore investigated other definitions of typical polymer
size. These include the {\it number average} $\ratn$, defined by
\begin{equation}
\ratn^2=\int d\sigp \sigp^2 \np(\sigp)
\label{eq:polhs_number_average}
\end{equation}
This is proportional to the average polymer chain length, which
scales as $\sigp^2$. The {\it mass average} $\ratm$ is also an average
chain length, but with the contribution from each polymer size
weighted by an additional factor of chain length (i.e.\ mass):
\begin{equation}
\ratm^2=\frac{\int d\sigp \sigp^4 \np(\sigp)}{\int d\sigp \sigp^2 \np(\sigp)}
\label{eq:polhs_mass_average}
\end{equation}
A final definition, the {\it virial average} $\ratb$, is derived from
the second virial coefficient for the polymer-colloid interaction. A
polymer's centre of mass is excluded from a sphere of diameter
$1+\sigp$ around a colloid of size $\sigc\equiv 1$, and $\ratb$ is
defined as the monodisperse polymer diameter giving the same average
excluded volume, viz.
\begin{equation}
(1+\ratb)^3=\int d\sigp (1+\sigp)^3 \np(\sigp)
\label{eq:polhs_virial_coeff}
\end{equation}
%
In a monodisperse system all of these definitions are of course
equivalent, so that $\rats=\ratn=\ratm=\ratb$. In a polydisperse
system they respond differently to the shape of the polymer diameter
distribution, and we will try to identify which definition is best
suited to characterising the phase behaviour in a manner that depends
only weakly on the polymer polydispersity. In assessing this
dependence we also have a choice for an appropriate measure of polymer
concentration. This could be e.g.\ the total polymer number density
$\rhp=\int\!d\sigp\,\rhp(\sigp)$ or the effective volume fraction of
the polymer spheres, $\phip=\int\!d\sigp \sigp^3 \rhp(\sigp)$;
values of $\phip$ of order unity
correspond to the crossover from the dilute to the semi-dilute polymer
regime. For a fixed diameter distribution $\np(\sigp)$, $\rhp$ and
$\phip$ are proportional to each other, but when comparing
distributions with different shapes the difference in the proportionality
coefficients matters. We choose mostly to work with $\phip$, following
Warren's suggestion~\cite{Warren97} that this produces phase diagrams
with a relatively weak dependence on the polymer diameter
distribution. This will allow us to assess how far Warren's insight
generalises beyond the binary mixtures of small polymers which he
considered.

As far as previous work on polymer polydispersity effects in
colloid-polymer mixtures goes, we have already mentioned Warren's work
for mixtures of polymers of two different sizes~\cite{Warren97}. This
explicitly accounted for the fact that the overall polymer density
distribution in the system -- which in the polydisperse case we will
write as $\rhp^{(0)}(\sigp)$ -- is imposed by the experimental
conditions. Studies of polymers with continuous size polydispersity,
on the other hand, have been restricted to the {\it
semi-grandcanonical} case~\cite{LeeRob99,SeaFre97}. (An exception is
very recent work~\cite{ParVarCumJac04} on interacting polymers; see
Sec.~\ref{sec:conclusion}.) Here the system is thought of as connected
to a large polymer reservoir which fixes the chemical potential
$\mupsig$ for each polymer size $\sigp$. The polymer density in the
system is determined only indirectly, and has to be deduced a
posteriori. Within the AO approach the polymers are treated as ideal
so that the reservoir density distribution and chemical potentials are
related by $\rhp^{\rm r}(\sigp)=e^{\mupsig}$. The decomposition of the
chemical potential into ideal and excess parts, $\mupsig = \ln\rhpsig
+ \mupex$ then shows that the polymer density distribution in any of
the phases within the system itself can be written as
\begin{equation}
\rhpsig = \rhp^{\rm r}(\sigp) e^{-\mupex}
\label{eq:polhs_map_s_c}
\end{equation}
Here the $\mupex$ are the excess polymer chemical potentials of the
phase being considered.


\begin{figure}
  \begin{center}
  \includegraphics[width=8cm]{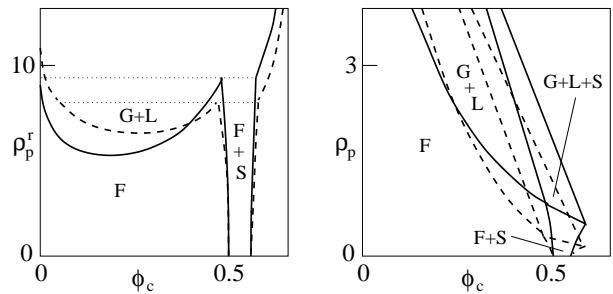}
  \caption{Phase diagram sketch of the phase behaviour for a
  semi-grandcanonical colloid-polymer mixture,
  following~\cite{SeaFre97}; F: fluid, G: gas, L: liquid, S:
  solid. The two graphs show the colloid volume fraction on the
  horizontal axis and on the vertical axis the polymer reservoir density
  $\rhp^{\rm r}$ (left) and the polymer density $\rhp$ in the system
  itself (right). The dashed and solid lines correspond to weak
  ($z\to\infty$, $\pol=0$) and
  strong ($z=5$) polymer polydispersity, respectively; the virial
  average polymer size is $\ratb=0.4$ in both cases.  The triple points
  are marked by horizontal dotted
  lines on the left; they are mapped to three-phase GLS triangles on the
  right.
\label{fig:pha_dia_pol_sketch}}
  \end{center}
\end{figure}
%
Fig.~\ref{fig:pha_dia_pol_sketch} shows a sketch of the results
obtained in Ref.~\cite{SeaFre97} for such a semi-grandcanonical
scenario, with polymer reservoir density distributions of Schulz form
(see Sec.~\ref{sec:polhs_Phase_behaviour} below). The graphs compare
the behaviour for weak ($z\to\infty$, i.e.\ $\pol=0$, dashed lines) and
strong ($z=5$, solid lines) polydispersity. The means of the Schulz
distributions were 
chosen so that the virial average polymer size $\ratb=0.4$ of the
normalised reservoir size distribution was the same in both cases. On
the left, the polymer reservoir density $\rhp^{\rm r}$ is sketched
against the colloid volume fraction.  Using\eq{eq:polhs_map_s_c},
these results can then be mapped onto a representation in terms of the
actual polymer density $\rhp$ in the system, as shown on the
right. This corresponds to fixing the total {\em number} of polymers
in the system while the polymer {\em size distribution} continues to
vary across the phase diagram, being controlled by the chemical
potentials -- or rather the chemical potential differences -- imposed
via the reservoir.

In both representations Fig.~\ref{fig:pha_dia_pol_sketch} shows
that, as a general rule, increasing the polymer polydispersity $\pol$
increases the size of the gas-liquid region in the phase diagram. As
seen most clearly in the plot on the right, however, there are also
regions in the phase diagram (at low $\rhp$) where gas-liquid
coexistence is suppressed with increasing $\pol$, and fluid-solid
phase separation is favoured. This seems counter-intuitive: size
polydispersity in the {\em colloids}, for example, is known to have
the opposite effect, delaying fluid-solid relative to gas-liquid phase
separation~\cite{FasSol03,FasSol04,FasSol04b}. One of our aims will be
to clarify this issue, and in particular the effect of the varying
polymer size distribution in the semi-grandcanonical setting.

The effects of polymer size polydispersity have also been analysed in
terms of the (pairwise part of the) depletion interaction between
colloids~\cite{Walz96,PieWal00,GouHan01,TuiPet02}. Conceptually, the
fixed polymer size distribution in these studies means again that
constant imposed polymer chemical potential are being considered; see
e.g. Ref.~\cite{LekPooPusStrWar92,DijVan97}. Finally, we mention the
related work of Ref.~\cite{DenSch02}. This also used a
semi-grandcanonical framework, but starting from a different physical
motivation: polymers were regarded as monodisperse with regard to chain
length, and size polydispersity in the equivalent polymer spheres was
considered as arising from compression of the polymers. In this case
the semi-grandcanonical approach is the physically relevant one since
polymer spheres of different size can be transformed into each
other. (A similar scenario arises when the polymers are replaced by
wormlike micelles~\cite{Poon02}.) The effects of polymer
compressibility on fluid-solid phase 
boundaries were found to be rather small up to size ratios
$\rat\approx 0.5$~\cite{DenSch02}. We therefore ignore them here and
focus on size polydispersity arising from a distribution of polymer
chain lengths. The experimental situation is then normally the canonical one,
where the overall distribution of polymer sizes in the system is
fixed.


Below we describe briefly the free energy which we use to model our
colloid-polymer mixture, and the numerical method by which phase
diagrams are obtained (Sec.~\ref{sec:polhs_Numerical_method}). Our
results for the effects of polydispersity on the phase behaviour are
given in Sec.~\ref{sec:polhs_Phase_behaviour}, separately for the
experimentally relevant case of fixed polymer size distribution and
for the semi-grandcanonical scenario where we compare with previous
studies~\cite{SeaFre97}. Conclusions and an outlook towards future
work can be found in the final section.

\section{Free energy and numerical method\label{sec:polhs_Numerical_method}}

In a recent paper~\cite{FasSol04b} we derived a free energy expression
for an AO system where both polymers and colloids are polydisperse, by
extending the free-volume approximation of Lekkerkerker {\em et
al}~\cite{LekPooPusStrWar92}. This free energy can be written in the
following form:
%
\begin{equation}
f = \fideal_{\rm c} + \fideal_{\rm p} + \fexhs
+ \int d\sigp\, \rhpsig \muex\hs(\sigp)
\label{eq:polhs_fre_ene_mix}
\end{equation}
The first two contributions are the ideal free energies of a mixture
of polydisperse colloids and polymers, respectively. The next term,
$\fexhs$, is the excess free energy of a system of pure (polydisperse)
hard-sphere colloids. In the last term, which represents the
colloid-polymer interaction, $\muex\hs(\sigp)$ is the excess chemical
potential of a hard-sphere particle of diameter $\sigp$ in the pure
colloid system.

Evaluation of the free energy~(\ref{eq:polhs_fre_ene_mix}) requires as
input only the properties of the pure colloid system, i.e.\ its excess
free energy $\fexhs$.
We therefore need to assign appropriate expressions for $\fexhs$ in
the colloidal fluid (or gas/liquid) and solid phases. For the {\it
fluid} part of the excess free energy the most accurate approximation
available is the BMCSL equation of
state~\cite{Boublik70,ManCarStaLel71} while for the {\it solid} we
adopt Bartlett's fit to simulation data for bidisperse hard sphere
mixtures~\cite{Bartlett99,Bartlett97}. Our previous
work~\cite{FasSol03,FasSol04} on polydisperse hard spheres has shown
that with these free energy expressions quantitatively accurate fits
to simulation data are obtained.
Both free energy expressions depend only on the moments $\rh_{{\rm
c}i}=\int d\sigc\rhcsig\sigc^i$ ($i=0\ldots 3$) of the colloid density
distribution. As a consequence the excess chemical potentials become
third order polynomials,
\begin{equation}
\muex\hs(\sig) = \frac{\delta\fexhs}{\delta\rhc(\sig)} = 
\sum_i \muex_{{\rm hs},i} \sig^i
\label{eq:muex_polynomial}
\end{equation}
where the $\muex_{{\rm hs},i}$ are the excess moment chemical
potentials of the pure hard-sphere system, $\muex_{{\rm hs},i} =
{\partial\fexhs} / {\partial\rhc{}_i}$. (As explained in
Ref.~\cite{FasSol04b}, we always evaluate these from the BMCSL free
energy.) The interaction term in\eq{eq:polhs_fre_ene_mix} then
simplifies to
\begin{equation}
\int d\sigp\, \rhpsig \muex\hs(\sigp) = \sum_i \muex_{{\rm hs},i}
\rh_{{\rm p}i}
\end{equation}
where the $\rh_{{\rm p}i}=\int d\sigp\rhpsig \sigp^i$ ($i=0\ldots 3$)
are the moments of the polymer density distribution.

We will deal with the particular case where the colloids are assumed
to be monodisperse. With the colloid diameter set to unity as assumed,
all colloid density moments then become identical, $\rh_{{\rm
c}i}=\rhc$. Writing out the ideal contributions explicitly, the free
energy\eq{eq:polhs_fre_ene_mix} thus takes the form
\begin{eqnarray}
f &=& \int d\sigp\, \rhpsig [\ln \rhpsig-1] + \rhc (\ln\rhc-1)
\nonumber \\
& & {}+{} \fexhs(\rhc) + \sum_i \muex_{{\rm hs},i}(\rhc) \mpi
\label{eq:polhs_f_monopol}
\end{eqnarray}
Here we have highlighted that $\fexhs$ (and therefore the $\muex_{{\rm
hs},i}$) now only depend on the colloid density $\rhc$, or
equivalently the colloidal volume fraction $\phic$; for monodisperse
colloids and in our units the two quantities are identical.  The
excess part of the free energy~(\ref{eq:polhs_f_monopol}) thus only
depends on the colloid density $\rhc$ and the moments $\rh_{{\rm p}i}$
of the polymer density distribution. As in~\cite{FasSol04b} for the
converse situation of polydisperse colloids and monodisperse polymers,
we can regard these quantities as moments of an enlarged density
distribution $(\rhc, \rhpsig)$. Because only a finite number (five) of
such moments are involved, we thus have a {\em truncatable} free
energy~\cite{SolWarCat01}. The phase equilibrium conditions associated
with this free energy can then be solved using the moment free energy
(MFE) method~\cite{SolCat98,Warren98,SolWarCat01,Sollich02,FasSol04b}. The
moment free energy allows one to map the full free
energy~(\ref{eq:polhs_f_monopol}), with its dependence on all details
of $\rhpsig$ through the ideal part, onto a MFE depending only on the
moments $\rhp{}_i$ and $\rhc$. From the latter, phase behaviour can
then be found by the conventional methods for finite mixtures,
treating each of the $\rhp{}_i$ as a number density of an appropriate
``quasi-species''~\cite{SolWarCat01}.
For truncatable free energies this locates exactly the cloud points,
i.e.\ the onset of phase separation coming from a single phase, as
well as the properties of the coexisting ``shadow'' phases that
appear.
Inside the coexistence region, one in principle needs to solve a set
of highly coupled nonlinear equations, and the predictions derived from
the MFE are only approximate. However, by retaining extra moments,
increasingly accurate solutions can be obtained by
iteration~\cite{SolWarCat01,ClaCueSeaSolSpe00,SpeSol02,SpeSol03a}.
Using these as initial points, we are then able to find the, for our free
energy, exact solutions of the phase equilibrium equations.
%
%
%

The MFE method is designed for the physically realistic scenario where
the overall polymer density distribution is conserved when the system
separates into two or more phases.
However, it
can also be exploited to analyse semi-grandcanonical scenarios.
%
%
This is done by relaxing the particle conservation constraints on some
moments; in our specific case, only $\rhc$ and $\rhp{}_0$ are kept
conserved while the other moments of the polymer density distribution
are adapted to minimise the free energy. One can show that this
reproduces the semi-grandcanonical analysis of Sear and
Frenkel~\cite{SeaFre97}, in full analogy to our discussion in
Ref.~\cite{FasSol04}.

\section{\label{sec:polhs_Phase_behaviour}Phase behaviour}

In this section we will describe our results for the overall phase
behaviour of a mixture of polydisperse polymers and monodisperse
hard-sphere colloids. Our numerical work requires a choice to be made
for the ``parent'' polymer diameter distribution $\np^{(0)}(\sigp)$
which is imposed when the system is prepared. We concentrate primarily
on a Schulz distribution
\[
\np^{(0)}(\sigp) \propto 
\sigp^z\exp\left[-\left(\frac{z+1}{\bar{\sig}}\right)\sigp\right].
\]
An upper cutoff of $\sigp=1$ on polymer sizes is imposed because for
larger sizes the chain structure of the polymers becomes important and
polymers can no longer be treated as effective spheres; see e.g.\
Refs.~\cite{Sear01,SchFuc02}.  The parameters $\bar{\sig}$ and $z$
control the mean and width of the Schulz distribution. Without a
cutoff, the mean size is simply $\rats=\bar{\sig}$ while the
polydispersity is related to $z$ via $\pol^2={1}/({z+1})$. In the
presence of the cutoff these relations remain valid for large $z$,
where the system is almost monodisperse; otherwise $z$ and $\bar\sig$
have to be calculated numerically to give the desired values of
$\rats$ and $\pol$. We will focus in our numerical work on three
values of the polydispersity, namely $\pol=1/\sqrt{20+1}\approx 0.22$,
$\pol=1/\sqrt{5+1}\approx 0.41$ and $\pol=1/\sqrt{2+1}\approx 0.58$.

To check the generality of our results, calculations have also been
performed for a triangular size distribution:
\[
\np^{(0)}(\sigp) = \frac{1}{w^2} \left\{
\begin{array}{lll}
\sigp-(\bar{\sig}-w) \quad & \mbox{for} & \bar{\sig}-w \leq \sigp \leq \bar{\sig} \\
(\bar{\sig}+w)-\sigp \quad & \mbox{for} & \bar{\sig} \leq \sigp \leq \bar{\sig}+w
\end{array}\right.
\]
Here the mean size is $\rats\equiv\bar{\sig}$ independently of the
width parameter $w$, which is related to the polydispersity by
$w=\sqrt{6}\,\bar{\sig}\pol$. We find that qualitative features do not
depend strongly on the shape of the polymer size distribution and so
mostly report only the results for a Schulz distribution.

\subsection{Phase diagrams\label{sec:polhs_Phase_diagram}}



%
\begin{figure}
  \begin{center}
\includegraphics[width=8cm]{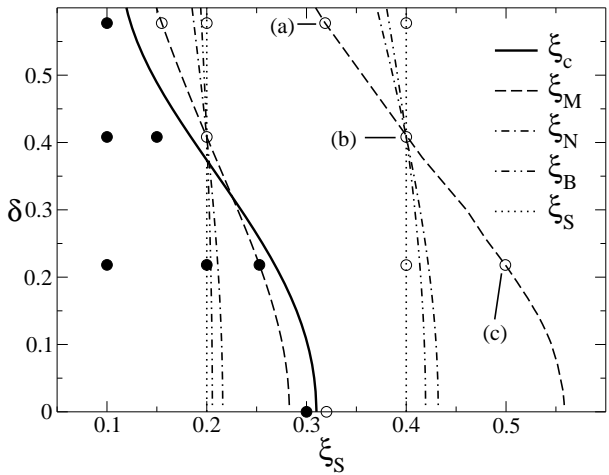} 
  \caption{Phase diagram topologies as a function of mean polymer size
  $\rats$ and polydispersity $\pol$. Full circles indicate parameter
  values where only fluid-solid coexistence is found, while empty
  circles show that the phase diagram has a gas-liquid coexistence
  region. The thick solid line shows a rough estimate for the
  boundary between the two regions, which defines the crossover value
  of the mean polymer size, $\ratc(\pol)$. Two sample sets of curves
  corresponding to constant $\rats$, $\ratn$, $\ratm$ and $\ratb$
  (chosen to agree at $\pol=0.41$) are also
drawn. Comparison 
with the thick solid line shows that $\ratm$ is most suitable for
predicting the phase diagram topology independently of $\pol$.
\label{fig:polhs_size_ratio}}
  \end{center}
\end{figure}
As discussed in the introduction, the phase diagrams of monodisperse
colloid-poymer mixtures may show either only fluid-solid coexistence
or, for sufficiently large polymers, an additional gas-liquid region
with an associated gas-liquid-solid triangle. We begin by
investigating the effect of polymer polydispersity on the change in
phase diagram topology as polymer size is increased, by calculating
phase diagrams for a range of values of mean polymer size $\rats$ and
polymer polydispersity $\pol$.  Fig.~\ref{fig:polhs_size_ratio}
summarises the results, for Schulz size distributions. The empty
circles indicate phase diagrams containing gas-liquid coexistence
regions, while the full circles identify phase diagrams where only a
fluid to fluid-solid transition was found. From these results we can
estimate the threshold value $\ratc$ for the mean polymer diameter as
a function of polydispersity $\pol$, as indicated by the thick solid
line in Fig.~\ref{fig:polhs_size_ratio}. The data show that as
polydispersity increases, $\ratc$ diminishes significantly. This is
due to the fact that, for given $\rats$, polydispersity tends to
favour gas-liquid over fluid-solid coexistence. At $\rats=0.2$, for
example, the phase diagram topology changes at $\pol\approx0.35$, from
fluid-solid coexistence only at low $\pol$ to additional gas-liquid
and gas-liquid-solid coexistence at high $\pol$.

\begin{figure}
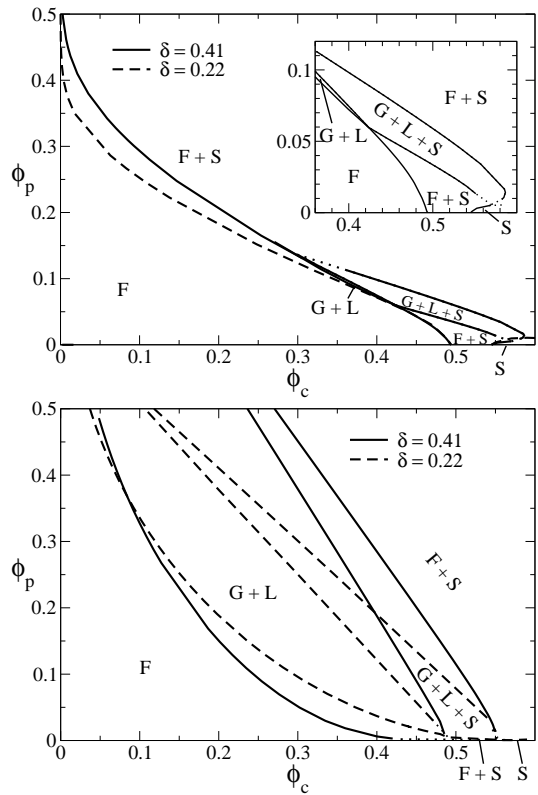

  \begin{center}
\includegraphics[width=7cm,clip=true]{./s_average_schulz_xi02_3.eps}  
\includegraphics[width=7cm,clip=true]{./s_average_schulz_xi04_3.eps}  
 \caption{Phase diagrams for Schulz distributions with
 polydispersity $\pol=0.41$ and $\pol=0.58$ and
 fixed mean polymer size $\rats=0.2$ (top) and $\rats=0.4$ (bottom),
 plotted in terms of effective polymer volume fraction $\phip$ versus
 colloid volume fraction $\phic$. Dotted line: best guess for the
 phase boundary where our numerical data become unreliable. The inset
 shows an enlarged view of the three-phase region for
 $\pol=0.41$ and $\rats=0.2$.
\label{fig:polhs_cloud_schulz}}
  \end{center}
\end{figure}
In Fig.~\ref{fig:polhs_cloud_schulz} we show the full phase diagram
obtained for Schulz distributions of polydispersity
$\pol=0.41$ and $\pol=0.58$ and at constant mean polymer
size of $\rats=0.2$ (top) and $\rats=0.4$ (bottom). The axis variables
are the colloid volume fraction $\phic$ -- identical in our units to
$\rhc$ -- and the (effective) polymer volume fraction
$\phip=\rhp{}_3$, following the suggestion by Warren~\cite{Warren97}
that the representation in terms of $\phip$ minimises polydispersity
effects on the phase behaviour. At $\rats=0.2$ we see explicitly how
gas-liquid and three-phase regions appear as $\pol$ is increased,
while the fluid-solid boundary recedes.
%
%
At $\rat=0.4$, Fig.~\ref{fig:polhs_cloud_schulz} (bottom), a
gas-liquid region is present for all polydispersities. The
main effect of increasing polydispersity is again to favour gas-liquid
coexistence. We note that, in contrast to what is found in a
monodisperse system, the phase boundaries of the gas-liquid-solid
region are no longer straight. This is due to the fact that in a
polydisperse system the constraint of particle conservation must be
satisfied for infinitely many different polymer sizes; indeed the
polymer size
distributions of the three coexisting phases changes as one moves
around the three-phase region.


%
%

We next ask which measure of effective polymer size is most appropriate
for predicting the phase diagram topology. Ideally, as explained in
the introduction, the value of this quantity should be enough to
predict the occurrence or not of gas-liquid coexistence, whatever the
polydispersity. In terms of Fig.~\ref{fig:polhs_size_ratio}, this
means that the curve $\ratc(\pol)$ should follow a contour of constant
effective polymer size. We have therefore plotted such contours for
each of the four possible measures of effective size discussed in the
introduction ($\rats$: mean diameter, $\ratn$: number average,
$\ratm$: mass average, $\ratb$: virial average). Two sets of contours
are shown, chosen such that all four quantities have identical values
($0.2$ and $0.4$, respectively) at polydispersity
$\pol=1/\sqrt{5+1}\approx 0.41$. This value of $\pol$ was also used as a
reference point in~\cite{SeaFre97}. We see that among the four
candidate definitions of an effective polymer size, the mass average
gives a contour which follows most closely -- although certainly not
perfectly -- the curve $\ratc(\pol)$ which marks the crossover between
the two phase diagram topologies. It is therefore our best candidate
for a useful measure of effective polymer size.

\begin{figure}
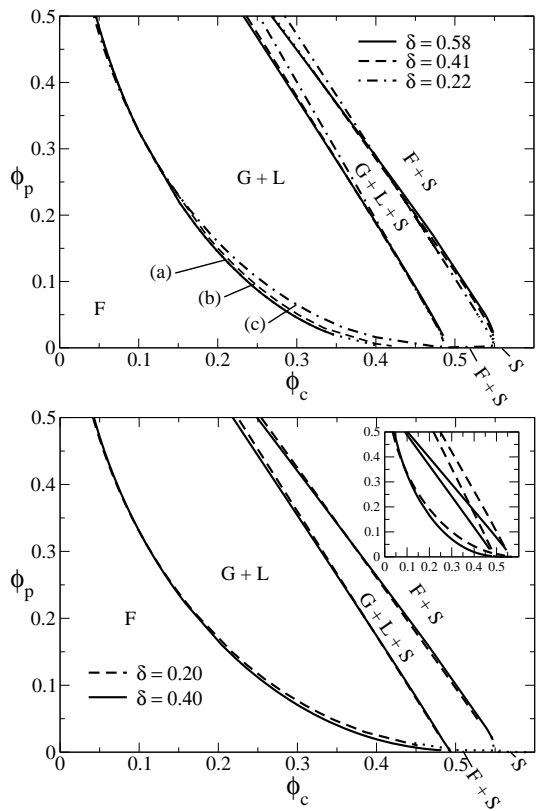

  \begin{center}
\includegraphics[width=7cm,clip=true]{./m_average_schulz_xi04_3.eps}
\includegraphics[width=7cm,clip=true]{./m_average_xi04_triangular2.eps} 
 \caption{Top: Phase diagrams for Schulz distributions with
 polydispersity $\pol=0.22$, 0.41 and 0.58 and constant mass
average $\ratm=0.558$; this should 
 be contrasted with Fig.~\ref{fig:polhs_cloud_schulz}, where the mean
 size $\rats$ ($=0.4$) was held fixed.
%
%
 Bottom: Phase diagrams for a triangular size distribution with
 $\pol=0.2$ and $0.4$ at constant mass average $\ratm=0.528$. In the
 inset the results for fixed mean diameter of $\rats=0.4$ are shown
 for comparison. Dotted lines indicate best guesses for phase
 boundaries we could not determine accurately from our numerics.
\label{fig:polhs_m_cloud_schulz}}
  \end{center}
\end{figure}
We can now go further and investigate whether polymer size
distributions with the same mass-average $\ratm$ have not only the
same phase diagram topology but in fact {\em quantitatively} similar
phase diagrams. As pointed out in the introduction, the answer to this
question also depends on how we represent the concentration of
polymers; we continue to use $\phip$. With this choice, the collapse
of phase diagrams for polymer size distributions with the same $\ratm$
is in fact rather good, as shown in
Fig.~\ref{fig:polhs_m_cloud_schulz} (top). The phase diagrams
displayed correspond to the parameter values at the points (a)-(c) in
Fig.~\ref{fig:polhs_size_ratio}. In
Fig.~\ref{fig:polhs_m_cloud_schulz} (bottom) we show analogous results
for triangular size distributions. Again the comparison at constant
mass-average $\ratm$ gives a good collapse of the phase diagrams for
different polydispersities, while at fixed $\rats$ polydispersity
causes significant changes; see the inset of
Fig.~\ref{fig:polhs_m_cloud_schulz} (bottom). In summary, we can say
that systems with the same 
mass-average $\ratm$ give phase diagrams which are not only
qualitatively but in fact quantitatively largely independent of
polydispersity. Our results suggest that this conclusion holds
independently of the polymer size distribution -- though multimodal
distributions might be expected to behave differently -- and up to
fairly substantial polydispersities of at least $40\%$ and possibly
higher.

We note finally that the results in
Fig.~\ref{fig:polhs_m_cloud_schulz} are for the case of relatively
large polymers, where phase separation is initially into gas and
liquid except at very low polymer concentrations. We have checked the
near $\pol$-independence of phase diagrams with the same $\ratm$ also
for smaller polymers, where fluid-solid phase separation occurs first
(data not shown), and reached the same conclusion. Very close to the
crossover in the phase diagram topology, deviations do of course occur.
This is clear from Fig.~\ref{fig:polhs_size_ratio}, which shows that
the contour of constant $\ratm$ does not follow the crossover curve
perfectly. In a narrow range of fixed values of $\ratm$ the phase
diagram will therefore change with $\pol$, both in terms of its
topology and the quantitative location of the phase boundaries.

\subsection{Comparison with semi-grandcanonical approach}


The qualitative trend of the results presented so far is that, for
fixed mean polymer size $\rats$, polymer polydispersity favours
gas-liquid coexistence over fluid-solid phase splits. On the other
hand, the results obtained by Sear and Frenkel in the
semi-grandcanonical approach~\cite{SeaFre97}, as sketched in
Fig.~\ref{fig:pha_dia_pol_sketch}, suggested that fluid-solid phase
separation can be favoured by polydispersity in some regions of the
phase diagram. This difference clearly needs to be understood.

Three possible causes suggest themselves. First, the results of Sear and
Frenkel were obtained by comparing reservoir polymer size
distributions with constant virial average size $\ratb$, whereas we
initially focused on constant $\rats$. Second, their phase diagrams
are represented in terms of polymer density $\rhp$ rather than effective
polymer volume fraction $\phip$. Third, their polymer size
distribution in the system varies because it is imposed only
indirectly via the reservoir size distribution, whereas it is fixed in
our analysis. 

\begin{figure}
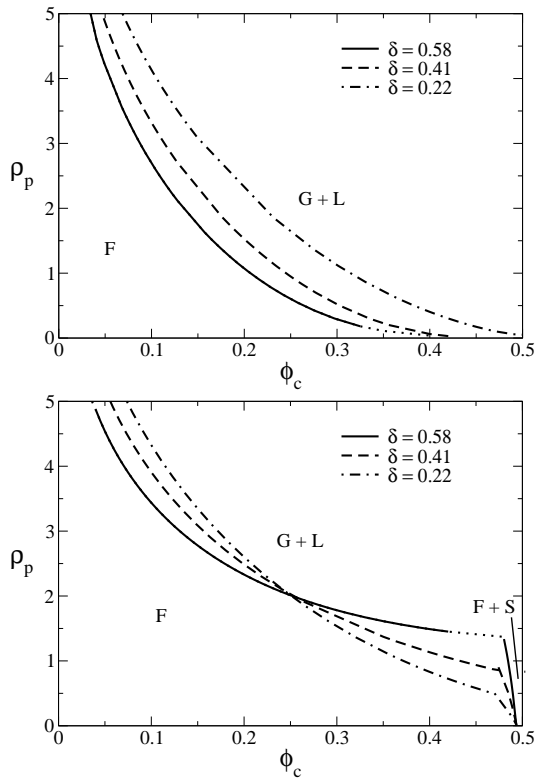

  \begin{center}
\includegraphics[width=7cm,clip=true]{./b2_xi04_3.eps} 
\includegraphics[width=7cm,clip=true]{./b2_comparison3.eps} 
 \caption{Cloud curves showing the onset of phase coexistence coming
  from low colloid density. Top: Canonical scenario; results are
  shown for three parent size distributions
  $\np^{(0)}(\sigp)$ with different polydispersities, all with the
  same virial average polymer size $\ratb$ as the reference
  distribution with $\rats=0.4$ and $\pol=0.41$. Bottom:
  Results for the corresponding semi-grandcanonical scenario, where
  the same three distributions are used as the {\em reservoir} size
  distribution, $\np^{\rm r}(\sigp)$.
\label{fig:polhs_semigrand}}
  \end{center}
\end{figure}
It is intuitively clear from the discussion above that keeping $\ratb$
rather than $\rats$ constant should not influence the results
greatly. Indeed, if we look at Fig.~\ref{fig:polhs_size_ratio} we see
that contours of constant $\ratb$ are close to those for constant
$\rats$, which are vertical lines. To rule out more quantitatively the
first two possible reasons discussed above, we have calculated the
cloud curves for 
the onset of gas-liquid coexistence for three Schulz size
distributions with identical $\ratb$, and plotted them in
Fig.~\ref{fig:polhs_semigrand} (top) with $\rhp$ on the vertical
axis. As before, gas-liquid coexistence is favoured by increased
polymer polydispersity. This leaves only the third reason, i.e.\ the
difference between the canonical and semi-grandcanonical
scenarios. Fig.~\ref{fig:polhs_semigrand} (bottom) compares our results
for the semi-grandcanonical scenario with the canonical ones in the
top graph. The same three polymer size distributions are used as before,
but they now specify the properties of the reservoir: the normalised
reservoir size distribution, $\np^{\rm r}(\sigp)$, is the same as the
parent size distribution $\np^{(0)}(\sigp)$ used in the canonical
case. As expected, our semi-grandcanonical results are in good
agreement~\cite{scaling} with those of Ref.~\cite{SeaFre97}. In
particular, we see in
Fig.~\ref{fig:polhs_semigrand} (bottom) that the change to the
semi-grandcanonical description has {\em reversed} the order of the
cloud curves at low polymer density as compared to the canonical
scenario shown in Fig.~\ref{fig:polhs_semigrand} (top).

We can thus conclude that the opposite trends with polydispersity seen
in the present study and in Ref.~\cite{SeaFre97} arise from the choice
of a semi-grandcanonical scenario in the latter. To understand
explicitly how the polymer size distribution is affected by this, we
can combine\eq{eq:polhs_map_s_c} and\eq{eq:muex_polynomial} to write
\begin{equation}
\rhp(\sigp) = \rhp^{\rm r}(\sigp)
\exp\left( - \sum_{i=0}^3 \muex_{{\rm hs},i} \sigp^i \right)
\label{eq:semigrand_mfe}
\end{equation}
%
where $\rhp^{\rm r}(\sigp)$ is the polymer density distribution in the
reservoir as before. The excess chemical potentials $\muex_{{\rm
hs},i}$ that appear here vary as we change the colloid density of our
system. Correspondingly the polymer density distribution has its shape
modified as we move around the phase diagram. In particular, at the
onset of phase coexistence the normalised size distribution
$\np(\sigp)$ will differ from the one in the reservoir, $\np^{\rm
r}(\sigp)$. In the more realistic canonical description, on the other
hand, $\np(\sigp)=\np^{(0)}(\sigp)$ remains fixed up to the point where
phase coexistence begins.

\begin{figure}
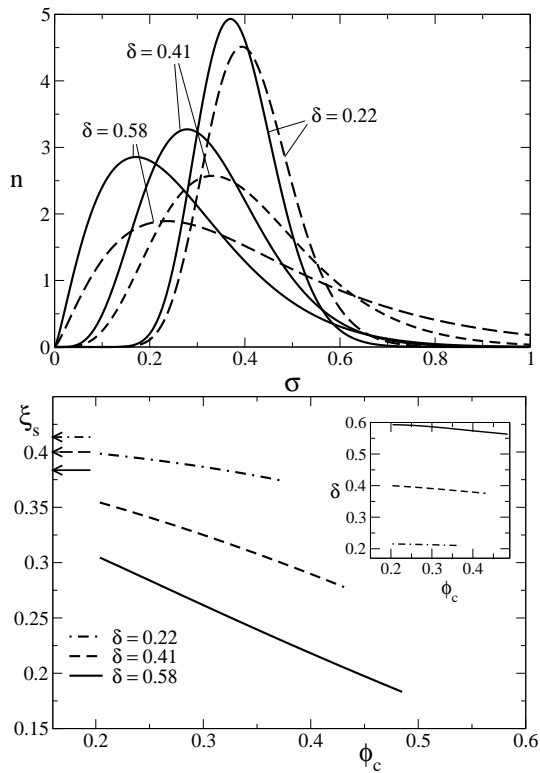

  \begin{center}
\includegraphics[width=7cm,clip=true]{./semigrand_distributions3.eps} 

\hspace*{1mm}\includegraphics[width=7cm,clip=true]{./parent_polydispersity_r3.eps} 
 \caption{Top: Comparison of polymer size distributions in a
 semi-grandcanonical system, $\np(\sigp)$ (solid lines), with the
 corresponding reservoir size distributions $\np^{\rm r}(\sigp)$
 (dashed).  The same three reservoir distributions as in
 Fig.~\ref{fig:polhs_semigrand} (bottom) are used; the colloid and
 polymer densities are fixed to $(\phic,\rhp)=(0.3,1.0)$. Bottom: Mean
 polymer size $\rats$ (main graph) and polydispersity $\pol$ (inset)
 of the system size distribution $\np(\sigp)$, at fixed polymer
 density $\rhp=1.0$ and for varying colloid density $\phic$. Arrows on
 the vertical axis in the main plot indicate the mean sizes in the
 reservoir.
\label{fig:polhs_semigrand_distribution}}
\end{center}
\end{figure}

%
To illustrate this difference, we plot in
Fig.~\ref{fig:polhs_semigrand_distribution} (top) the normalised size
distributions $\np(\sigp)$ in the semi-grandcanonical system (solid
lines). These are compared to the reservoir size distributions
$\np^{\rm r}(\sigp)$ (dashed lines), which are the same as used in
Fig.~\ref{fig:polhs_semigrand} (bottom); the colloid and polymer
densities are $(\phic,\rhp)=(0.3,1.0)$. We notice that the system and
reservoir size distributions differ significantly: the Gibbs-Boltzmann
factor in\eq{eq:semigrand_mfe} shifts the mean polymer size to lower
values, and this effect becomes stronger as the polydispersity
increases. Fig.~\ref{fig:polhs_semigrand_distribution} (bottom) traces
the mean and polydispersity of the system size distribution as a
function of colloid volume fraction $\phic$, at fixed $\rhp=1.0$.  The
reduction in the mean size compared to the reservoir is seen to
increase with colloid density. The polydispersity, on the other hand,
is only weakly affected, as the inset of
Fig.~\ref{fig:polhs_semigrand_distribution} (bottom) shows.

The lines in Fig.~\ref{fig:polhs_semigrand_distribution} (bottom) are
drawn up to the cloud point where phase separation begins. For the
widest reservoir distribution ($\pol=0.58$), the mean polymer
size in the system at this point is roughly half that in the
reservoir. Because smaller polymers produce a depletion interaction
that is both weaker and more short-ranged, this reduction in polymer
size delays the onset of phase separation. Overall, we conclude that
the inversion in the order of the cloud curves in the bottom part of
Fig.~\ref{fig:polhs_semigrand} (bottom) results from the shift of the
polymer size distribution to lower values as the polydispersity of the
reservoir is increased.

The reduction of the polymer mean size in the semi-grandcanonical
scenario can be understood quantitatively for narrow reservoir
distributions. Looking at\eq{eq:semigrand_mfe}, if $\rhp^{\rm
r}(\sigp)$ is narrowly peaked around $\sigp=\rats^{\rm r}$, then we
can expand the exponential as
$\mbox{const}\times[1-\lambda(\sigp-\rats^{\rm r})]$ to leading
order. From this one easily sees that the mean size $\rats$ in the system
is
\begin{equation}
\rats = \rats^{\rm r}-\lambda\langle(\sigp-\rats^{\rm
r})^2\rangle_{\rm r} = \rats^{\rm r}-\lambda\pol^2(\rats^{\rm r})^2
\label{eq:rats_expansion}
\end{equation}
to lowest order in $\pol$. The coefficient $\lambda$ must be positive
because the excess chemical potential in a hard-sphere system,
$\muex_{\rm hs}(\sigp)$, is an increasing function of $\sigp$. As
expected we have, therefore, a reduction in the mean polymer size
compared to the reservoir, by an amount which grows with the reservoir
polydispersity as $\pol^2$.


Finally, we note that the three curves in
Fig.~\ref{fig:polhs_semigrand} (bottom) all appear to cross at the
same point. To see how this effect arises, we focus again on narrow
reservoir distributions. The location of the cloud point value of the
polymer density $\rhp=\rhp(\phic,\rats,\pol)$ depends on the colloid
volume fraction $\phic$, the mean polymer size $\rats$ in the system
and the polydispersity $\pol$; for small $\pol$, other properties of
the size distribution are irrelevant. In a semi-grandcanonical
setting, $\rats$ and $\pol$ are determined indirectly by the colloid
volume fraction $\phic$ as well as the mean polymer size $\rats^{\rm
r}$ and the polydispersity of the reservoir. In a perturbation
expansion to order $\pol^2$ we can approximate $\pol^2$ by its fixed
value in the reservoir, and write $\rats = \rats^{\rm r} -
\lambda\pol^2(\rats^{\rm r})^2$ from\eq{eq:rats_expansion}. The
explicit dependence on $\pol$ of the cloud point is also expected to
be quadratic to leading order~\cite{EvaFaiPoo98,Evans01}. Overall, we
can therefore expand the cloud point position in a weakly polydisperse
semi-grandcanonical system as
\begin{eqnarray}
\rhp(\phic,\rats,\pol) &=& \rhp(\phic,\rats^{\rm r},0) -
\lambda\pol^2(\rats^{\rm r})^2
\frac{\partial\rhp}{\partial\rats}(\phic,\rats^{\rm r},0) 
\nonumber \\
& & {}+{} \pol^2
\frac{\partial\rhp}{\partial\pol^2}(\phic,\rats^{\rm r},0) 
\label{eq:rhp_perturbation}
\end{eqnarray}
How this varies with increasing polydispersity $\pol^2$ depends on
the competition of the last two terms. The last term is also present
in the canonical scenario and represents the shift in the cloud point
at constant mean polymer size. We saw above that polydispersity
favours phase coexistence under these conditions, so that
$\partial\rhp/\partial\pol^2$ must be negative; see also
Fig.~\ref{fig:polhs_semigrand} (top). The second term is only present
for the semi-grandcanonical case; because increasing $\rats$ again
favours phase separation, we have that also
$\partial\rhp/\partial\rats$ is negative.
Thus the two $\pol^2$-terms in\eq{eq:rhp_perturbation} have opposite
signs. The coefficient $\lambda$ depends on $\phic$; it tends to zero
for $\phic\to 0$, where the excess chemical potentials in a
hard-sphere system vanish, and grows from there with increasing
$\phic$. At small $\phic$ we therefore expect the third term
in\eq{eq:rhp_perturbation} to dominate, so that polydispersity favours
phase separation as in the canonical case. As $\phic$ increases, the
second term eventually balances the third. At this point the cloud
curve is unaffected by $\pol^2$ to leading order, and this produces 
the crossing phenomenon in Fig.~\ref{fig:polhs_semigrand}
(bottom). At even higher $\phic$, the second term dominates and so
phase separation is delayed with increasing $\pol$, exactly as seen in
the lower part of Fig.~\ref{fig:polhs_semigrand} (bottom).

Overall, while in the realistic canonical description polydispersity
always favours gas-liquid phase separation, in the semi-grandcanonical
scenario this trend can be counteracted by a decrease in the mean
polymer size in the system. This effect can become dominant at high
colloid volume fractions, producing the reversal in the order of the
curves in the lower parts of Fig.~\ref{fig:polhs_semigrand} (top) and
(bottom).

\section{\label{sec:4hspolConclusionandoutlook}Conclusion and outlook}
\label{sec:conclusion}

We have investigated the effect of polymer polydispersity on a mixture
of ideal polymers and hard-sphere colloids. Our focus was on the
realistic (canonical) case where the distribution of polymer sizes is
fixed when the system is prepared, but we have also compared with
semi-grandcanonical approaches where the size distribution adapts by
equilibrating to chemical potentials imposed by a large polymer
reservoir~\cite{SeaFre97}. For the canonical setting, gas-liquid phase
separation is favoured by increased polydispersity while fluid-solid
phase separation is retarded. This is in broad agreement with the
intuition that polydispersity disfavours ordered
phases~\cite{FasSol04,FasSol04b}, though less obvious here since the
size polydispersity is in the polymers, not in the colloids that
order translationally in the solid.

In a semi-grandcanonical scenario the above trends with polydispersity
are reversed in some parts of the phase diagram~\cite{SeaFre97}. We
saw that this arises because the polymer size distribution in the
system can become rather different from the reservoir size
distribution, especially for dense colloidal phases.  Specifically,
the size distribution is shifted to smaller polymer sizes, the more so
the larger the degree of polydispersity.  This effect is sufficiently
strong to reverse the order of the cloud curves compared to the
canonical setting, favouring fluid-solid over gas-liquid formation.

For the canonical case of conserved polymer size distribution, we
established that the phase diagram topology was nearly independent of
polydispersity provided one compares systems with the same mass
average polymer size $\ratm$. We were lead to this observation by
estimating the location of the curve $\ratc(\pol)$ in the
$(\rats,\pol)$ plane along which the phase diagram topology changes,
and comparing this with contours of constant $\ratm$. The crossover
value $\ratc$ decreases with increasing polydispersity $\pol$, showing
quantitatively that gas-liquid coexistence is favoured as
polydispersity is increased at constant mean polymer size. Finally, we
showed that even quantitatively the phase diagrams of systems with the
same mass average $\ratm$ are nearly independent of polydispersity,
provided that polymer concentration is represented in terms of the
effective polymer volume fraction $\phip$.

To assess the reliability of our results, we discuss briefly the
approximations made. We have employed the standard AO model instead of
explicitly modelling ideal polymer chains. Simulations~\cite{MeiFre94}
suggest that this is a reasonable approximation for polymers which are
not too large, i.e.\ $\sigp<1$. A slight improvement can be obtained
by using a mapping from polymer radius of gyration to effective sphere
radius $\sigp/2$ which accounts for the effect of the curvature of the
colloid surface on the polymer
conformations~\cite{MeiFre94,LouBolMeiHan02}. In the limit of a small
polymer, this results in an increase of $\sigp$ by a factor
$2/\sqrt{\pi}\approx 1.13$ over the naive value, 
but in the range
$0.2<\sigp<1$ of interest the corrections to $\sigp$ are small
($<5\%$). With such corrections included,
a Schulz distribution across the polymers' radii of gyration -- the
case we concentrated on -- would result in a distribution of $\sigp$
that is no longer precisely of Schulz form; quantitatively, however,
this will again be a small effect. 
We have also used the free-volume approximation to estimate
the free energy of the polydisperse AO model. As discussed
elsewhere~\cite{FasSol04b}, this is essentially equivalent to the
density functional theory approach of Ref.~\cite{DenSch02}; see also
the discussion for the monodisperse case in
Ref.~\cite{SchLowBraEva02}. Again, simulations suggest that this
approach is reliable~\cite{MeiFre94,BolLouHan02} in the range
$0.3<\sigp<1$, although somewhat less so after polymer reservoir
densities $\rhp^{\rm r}$ are converted back to those in the system
(compare e.g.\ Fig.\ 3 in Ref.~\cite{BolLouHan02} with Fig.\ 2(b) in
Ref.~\cite{SchLowBraEva02}).

We comment briefly on comparisons with experimental work. Much work on
mixtures of colloids and $\theta$-point polymers has been done in
Edinburgh; see e.g.\ the review~\cite{Poon02}. Experimentally, the
typical polymer size is normally determined from the weight-average
molecular weight, which corresponds to our $\ratm^2$. Interestingly, our
results then suggest that polymer polydispersity should have only a
small effect on the phase behaviour, so that e.g.\ the comparison
between experimental data and AO model predictions
in~\cite{IleOrrPooPus95} would remain essentially
unchanged~\cite{conc_note}. A systematic experimental study of such
polydispersity effects would seem worthwhile. For example, one could
prepare two polydisperse systems, one whose number-averaged molecular
weight ($\sim \ratn^2$) and one whose mass-averaged molecular weight
($\sim \ratm^2$) coincides with that of a near-monodisperse reference
system; our theory would then predict that the second polydisperse
system produces phase behaviour much more similar to the monodisperse
reference than the first. Also fruitful could be the investigation of
polymer size fractionation between coexisting phases. We have not
displayed our theoretical predictions for this; one finds as in the
semi-grandcanonical approach~\cite{SeaFre97} that the larger polymers
are found in the phases with lower colloid density, but the detailed
distribution shapes are different because they always need to combine
to give the fixed parent distribution.

An interesting extension of the present work would be to consider the
combined effect of polymer {\em and} colloid size polydispersity,
generalising the present study and our previous one on polydisperse
colloids and monodisperse polymers~\cite{FasSol04b}. This is in
principle possible using the current framework, but the excess free
energy would then depend on eight moments $\rhc{}_i$ and $\rhp{}_i$
($i=0\ldots 3$), making the problem very challenging
numerically. Physically, one would expect from our current results
that the crossover value of the polymer size, when measured in terms
of the mass average $\ratm$, would be largely unaffected by {\em
polymer} polydispersity, but decrease significantly with increasing
{\em colloid} polydispersity~\cite{FasSol04b}.
A simpler alternative to the treatment of a fully polydisperse system 
would be to consider the colloid size distribution
as ``quenched'', i.e.\ equal to the parent in all phases, thus
excluding colloid size fractionation. One would then need only one
conserved colloid moment in the excess free energy, and the complexity
of the numerical analysis would be the same as in the present
study. Physically, this quenched approximation could be appropriate
for describing the initial stages of phase separation in a
polydisperse colloid-polymer mixture, where {\em polymer} size
fractionation -- which should be the faster process -- has already
taken place while {\em colloid} size fractionation is still
negligible.

A key open question for further work is how our results generalize to
the case of good or poor (rather than $\theta$-) solvents, where
polymer-polymer interactions can no longer be neglected. For
monodisperse polymers this issue has been tackled by a variety of
methods including virial expansions near the
$\theta$-point~\cite{WarIlePoo95}, integral
equations~\cite{FucSch00,FucSch01,FucSch02,RamFucSchZuk02},
Flory-Huggins theory~\cite{Sear02}, coarse-graining of polymers into
soft colloids~\cite{BolLouHan02,LouBolMeiHan02,RotDzuHanLou04},
modifications of free-volume theory~\cite{AarTuiLek02}, and an AO model
with added soft polymer-polymer repulsions~\cite{SchDenBra03} to model
good solvents or with a repulsive co-solvent to mimic poor
ones~\cite{SchDen02}. It remains an open challenge to incorporate
polymer polydispersity into these approaches. Some progress in this
direction has already been made in Ref.~\cite{ParVarCumJac04}, which
was published after the present work was completed. In this study the
polymers were modelled as chains of hard-sphere particles, as would be
appropriate for good solvent conditions, and calculations for the
onset of gas-liquid coexistence were performed using a truncatable
model free energy derived from thermodynamic perturbation
theory~\cite{ParVarJac03}. Also in this scenario with interacting
polymers, it is found that polydispersity significantly enlarges the
size of the coexistence region. It appears therefore that this
qualitative conclusion is robust to the presence or otherwise of
polymer non-ideality. If the approach of Ref.~\cite{ParVarCumJac04}
can be extended to include colloidal solids, it would be interesting
to ask whether polydispersity effects on the relative stability of
fluid-solid and gas-liquid phase separation are likewise robust under
the inclusion of polymer-polymer interactions.

The authors acknowledge support of the EPSRC through grant number
GR/R52121/01.


\bibliography{/home/psollich/references/references,local}

\begin{thebibliography}{10}

\bibitem{Vincent74}
B. Vincent, Adv.\ Colloid Interface Sci. {\bf 4},  193  (1974).

\bibitem{Napper83}
D. Napper, {\em Polymeric stabilization of colloidal dispersions} (Academic,
  New York, 1983).

\bibitem{RusSavSch89}
W.~B. Russel, D.~A. Saville, and W.~R. Schowalter, {\em Colloidal dispersions}
  (Cambridge University Press, Cambridge, 1989).

\bibitem{EniFriStaPop90}
N.~S. Enikolopyan, M.~L. Fridman, I.~O. Stalnova, and V.~L. Popov, Adv.\
  Polym.\ Sci. {\bf 96},  1  (1990).

\bibitem{Lewis00}
J.~A. Lewis, J.\ Am.\ Ceram.\ Soc. {\bf 83},  2341  (2000).

\bibitem{AsaOos54}
S. Asakura and F. Oosawa, J.\ Chem.\ Phys. {\bf 22},  1255  (1954).

\bibitem{AsaOos58}
S. Asakura and F. Oosawa, J.\ Polym.\ Sci. {\bf 33},  183  (1958).

\bibitem{Vrij76}
A. Vrij, Pure Appl.\ Chem. {\bf 48},  471  (1976).

\bibitem{JoaLeiDeG79}
J.~F. Joanny, L. Leibler, and P.~G. {de Gennes}, J.\ Polym.\ Sci.\ B {\bf 17},
  1073  (1979).

\bibitem{MeiFre91}
E.~J. Meijer and D. Frenkel, Phys.\ Rev.\ Lett. {\bf 67},  1110  (1991).

\bibitem{MeiFre94}
E.~J. Meijer and D. Frenkel, J.\ Chem.\ Phys. {\bf 100},  6873  (1994).

\bibitem{GasHalRus83}
A.~P. Gast, C.~K. Hall, and W.~B. Russel, J.\ Colloid Interface Sci. {\bf 96},
  251  (1983).

\bibitem{Vincent87}
B. Vincent, Colloid Surface {\bf 24},  269  (1987).

\bibitem{VinEdwEmmCro88}
B. Vincent, J. Edwards, S. Emmett, and R. Croot, Colloid Surface {\bf 31},  267
   (1988).

\bibitem{LekPooPusStrWar92}
H.~N.~W. Lekkerkerker {\it et~al.}, Europhys.\ Lett. {\bf 20},  559  (1992).

\bibitem{SchLowBraEva00}
M. Schmidt, H. L{\"{o}}wen, J.~M. Brader, and R. Evans, Phys.\ Rev.\ Lett. {\bf
  85},  1934  (2000).

\bibitem{SchLowBraEva02}
M. Schmidt, H. L{\"{o}}wen, J.~N. Brader, and R. Evans, J.\ Phys.\ Cond.\ Matt.
  {\bf 14},  9353  (2002).

\bibitem{BraEvaSch03}
J.~M. Brader, R. Evans, and M. Schmidt, Mol.\ Phys. {\bf 101},  3349  (2003).

\bibitem{DijBraEva99}
M. Dijkstra, J.~M. Brader, and R. Evans, J.\ Phys.\ Cond.\ Matt. {\bf 11},
  10079  (1999).

\bibitem{BolLouHan02}
P.~G. Bolhuis, A.~A. Louis, and J.~P. Hansen, Phys.\ Rev.\ Lett. {\bf 89},
  128302  (2002).

\bibitem{FasSol04b}
M. Fasolo and P. Sollich, J.\ Chem.\ Phys.  (2004), to be published. Preprint
  cond-mat/0410374.

\bibitem{Warren97}
P.~B. Warren, Langmuir {\bf 13},  4588  (1997).

\bibitem{LeeRob99}
J.~T. Lee and M. Robert, Phys.\ Rev.\ E {\bf 60},  7198  (1999).

\bibitem{SeaFre97}
R.~P. Sear and D. Frenkel, Phys.\ Rev.\ E {\bf 55},  1677  (1997).

\bibitem{ParVarCumJac04}
P. Paricaud, S. Varga, P.~T. Cummings, and J. G, Chem.\ Phys.\ Lett. {\bf 398},
   489  (2004).

\bibitem{FasSol03}
M. Fasolo and P. Sollich, Phys.\ Rev.\ Lett. {\bf 91},  068301  (2003).

\bibitem{FasSol04}
M. Fasolo and P. Sollich, Phys.\ Rev.\ E {\bf 70},  041410  (2004).

\bibitem{Walz96}
J.~Y. Walz, J.\ Colloid Interface Sci. {\bf 178},  505  (1996).

\bibitem{PieWal00}
M. Piech and J.~Y. Walz, J.\ Colloid Interface Sci. {\bf 225},  134  (2000).

\bibitem{GouHan01}
D. Goulding and J.~P. Hansen, Mol.\ Phys. {\bf 99},  865  (2001).

\bibitem{TuiPet02}
R. Tuinier and A.~V. Petukhov, Macromol.\ Theory Simul. {\bf 11},  975  (2002).

\bibitem{DijVan97}
M. Dijkstra and R. {van Roij}, Phys.\ Rev.\ E {\bf 56},  5594  (1997).

\bibitem{DenSch02}
A.~R. Denton and M. Schmidt, J.\ Phys.\ Cond.\ Matt. {\bf 14},  12051  (2002).

\bibitem{Poon02}
W.~C.~K. Poon, J.\ Phys.\ Cond.\ Matt. {\bf 14},  R859  (2002).

\bibitem{Boublik70}
T. Boublik, J.\ Chem.\ Phys. {\bf 53},  471  (1970).

\bibitem{ManCarStaLel71}
G.~A. Mansoori, N.~F. Carnahan, K.~E. Starling, and T.~W. {Leland, Jr.}, J.\
  Chem.\ Phys. {\bf 54},  1523  (1971).

\bibitem{Bartlett99}
P. Bartlett, Mol.\ Phys. {\bf 97},  685  (1999).

\bibitem{Bartlett97}
P. Bartlett, J.\ Chem.\ Phys. {\bf 107},  188  (1997).

\bibitem{SolWarCat01}
P. Sollich, P.~B. Warren, and M.~E. Cates, Adv.\ Chem.\ Phys. {\bf 116},  265
  (2001).

\bibitem{SolCat98}
P. Sollich and M.~E. Cates, Phys.\ Rev.\ Lett. {\bf 80},  1365  (1998).

\bibitem{Warren98}
P.~B. Warren, Phys.\ Rev.\ Lett. {\bf 80},  1369  (1998).

\bibitem{Sollich02}
P. Sollich, J.\ Phys.\ Cond.\ Matt. {\bf 14},  R79  (2002).

\bibitem{ClaCueSeaSolSpe00}
N. Clarke {\it et~al.}, J.\ Chem.\ Phys. {\bf 113},  5817  (2000).

\bibitem{SpeSol02}
A. Speranza and P. Sollich, J.\ Chem.\ Phys. {\bf 117},  5421  (2002).

\bibitem{SpeSol03a}
A. Speranza and P. Sollich, J.\ Chem.\ Phys. {\bf 118},  5213  (2003).

\bibitem{Sear01}
R.~P. Sear, Phys.\ Rev.\ Lett. {\bf 86},  4696  (2001).

\bibitem{SchFuc02}
M. Schmidt and M. Fuchs, J.\ Chem.\ Phys. {\bf 117},  6308  (2002).

\bibitem{scaling}
It appears that Fig.\ 4 in~\protect\cite{SeaFre97} may have been incorrectly
  scaled: on the horizontal axis, which represents a pure hard-sphere colloid
  system without polymer, the onset of fluid-solid coexistence is shown at
  $\phic\approx0.525$ rather than the correct value $\phic=0.494$ which our
  calculation and the other figures in~\protect\cite{SeaFre97} reproduce.

\bibitem{EvaFaiPoo98}
R.~M.~L. Evans, D.~J. Fairhurst, and W.~C.~K. Poon, Phys.\ Rev.\ Lett. {\bf
  81},  1326  (1998).

\bibitem{Evans01}
R.~M.~L. Evans, J.\ Chem.\ Phys. {\bf 114},  1915  (2001).

\bibitem{LouBolMeiHan02}
A.~A. Louis, P.~G. Bolhuis, E.~J. Meijer, and J.~P. Hansen, J.\ Chem.\ Phys.
  {\bf 117},  1893  (2002).

\bibitem{IleOrrPooPus95}
S.~M. Ilett, A. Orrock, W.~C.~K. Poon, and P.~N. Pusey, Phys.\ Rev.\ E {\bf
  51},  1344  (1995).

\bibitem{conc_note}
However, care is needed when comparing measures of polymer concentrations.
  Whereas we have used the effective polymer volume fraction $\phip$,
  experimentally the polymer mass density is more readily accessible; in our
  units, the latter is proportional to $\rhp\ratn^2$. At constant $\ratm$, one
  finds e.g.\ for a Schulz distribution that the ratio
  $\phip/(\rhp\ratn^2)=\langle \sigp^3\rangle/\langle \sigp^2\rangle$ {\em
  decreases} with increasing polymer polydispersity $\pol$. Since we found
  phase diagrams that were relatively insensitive to $\pol$ when expressed in
  terms of $\phip$, the corresponding values of polymer mass density should
  {\em increase} with $\pol$. Quantitatively the effect is moderate, however,
  leading to an increase of only $\approx 10\%$ over the monodisperse case even
  for $\pol=0.6$.

\bibitem{WarIlePoo95}
P.~B. Warren, S.~M. Ilett, and W.~C.~K. Poon, Phys.\ Rev.\ E {\bf 52},  5205
  (1995).

\bibitem{FucSch00}
M. Fuchs and K.~S. Schweizer, Europhys.\ Lett. {\bf 51},  621  (2000).

\bibitem{FucSch01}
M. Fuchs and K.~S. Schweizer, Phys.\ Rev.\ E {\bf 64},  021514  (2001).

\bibitem{FucSch02}
M. Fuchs and K.~S. Schweizer, J.\ Phys.\ Cond.\ Matt. {\bf 14},  R239  (2002).

\bibitem{RamFucSchZuk02}
S. Ramakrishnan, M. Fuchs, K.~S. Schweizer, and C.~F. Zukoski, Langmuir {\bf
  18},  1082  (2002).

\bibitem{Sear02}
R.~P. Sear, Phys.\ Rev.\ E {\bf 66},  051401  (2002).

\bibitem{RotDzuHanLou04}
B. Rotenberg, J. Dzubiella, J.~P. Hansen, and A.~A. Louis, Mol.\ Phys. {\bf
  102},  1  (2004).

\bibitem{AarTuiLek02}
D.~G. A.~L. Aarts, R. Tuinier, and H.~N.~W. Lekkerkerker, J.\ Phys.\ Cond.\
  Matt. {\bf 14},  7551  (2002).

\bibitem{SchDenBra03}
M. Schmidt, A.~R. Denton, and J.~M. Brader, J.\ Chem.\ Phys. {\bf 118},  1541
  (2003).

\bibitem{SchDen02}
M. Schmidt and A.~R. Denton, Phys.\ Rev.\ E {\bf 65},  061410  (2002).

\bibitem{ParVarJac03}
P. Paricaud, S. Varga, and G. Jackson, J.\ Chem.\ Phys. {\bf 118},  8525
  (2003).

\end{thebibliography}
\bibliographystyle{prsty}

\end{document}